\documentclass[aps,twocolumn,superscriptaddress]{revtex4}
\usepackage{times,latexsym,amssymb}
\usepackage[dvips]{graphicx,color}

\newcommand{\ignore}[1]{}

\newcommand{\mComment}[1]{}
\newcommand{\rComment}[1]{}
\newcommand{\gComment}[1]{}
\renewcommand{\mComment}[1]{\textcolor{blue}{Manny: #1}}
\renewcommand{\rComment}[1]{\textcolor{magenta}{Ray: #1}}
\renewcommand{\gComment}[1]{\textcolor{red}{Gerard: #1}}
\newcommand{\ket}[1]{|{#1}\rangle}

\newcommand{\kets}[2]{|{#1}\rangle_{{}_{\!\!{#2}}}}

\newcommand{\slb}[2]{{{#1}^{({#2})}}}

\newcommand{\defeq}{:=}

\newlength{\isplength}
\newcounter{ispcount}
\newcommand{\ssp}{\setcounter{ispcount}{1}\setlength{\isplength}{\parindent}\hspace*{\isplength}}
\newcommand{\isp}{\addtocounter{ispcount}{1}\setlength{\isplength}{\value{ispcount}\parindent}\hspace*{\isplength}}

\begin{document}

\title{Thresholds for Linear Optics Quantum Computation}
\author{E. Knill}
\email[]{knill@lanl.gov}
\homepage[]{http://www.c3.lanl.gov:~knill}
\author{R. Laflamme}
\email[]{laflamme@lanl.gov}
\affiliation{Los Alamos National Laboratory, MS B265, Los Alamos, New Mexico 87545}
\author{G. Milburn}
\email[]{milburn@physics.uq.edu.au}
\affiliation{Center for Quantum Computer Technology, Department of Physics, University of Queensland, St. Lucia, Australia}

\begin{abstract}
We previously established that in principle, it is possible to quantum
compute using passive linear optics with
photo-detectors~\cite{knill:qc2000b}. Here we describe techniques
based on error detection and correction that greatly improve the
resource and device reliability requirements needed for
scalability. The resource requirements are analyzed for ideal linear
optics quantum computation (LOQC).  The coding methods can be
integrated both with loss detection and phase error-correction to deal
with the primary relaxation processes in non-ideal optics, including
detector inefficiencies. The main conclusion of our work is that the
resource requirements for implementing quantum communication or
computation with LOQC are reasonable.  Furthermore, this work clearly
demonstrates how special knowledge of the error behavior can be
exploited for greatly improving the fault tolerance and overheads of a
physical quantum computer.
\end{abstract}
%\pacs{03.67.Lx, 89.70.+c}

\maketitle

\section{Introduction}

In~\cite{knill:qc2000b} we proposed that linear optics quantum
computation (LOQC) is a viable option for physically realizing quantum
computers. The proposal depends on a series of optical protocols that
require single photon state preparation and measurement whose outcome
can be used to control other optical elements. The basic idea is to
kick back the hidden non-linearities in photo-detectors to qubits
encoded in pairs of optical modes (photonic qubits).  The protocols
were shown to implement the necessary quantum gates with arbitrarily
high probability of success. In their simplest form, the resources
required for implementing the protocols grow rapidly as the desired
success probability is increased.  We noted that due to the general
accuracy threshold
theorem~\cite{aharonov:qc1996a,kitaev:qc1997a,knill:qc1998a,preskill:qc1998a},
the model scales efficiently asymptotically, with constant overhead
for implementing the standard fault tolerant model of quantum computation.
We also suggested that with the use of erasure
codes~\cite{grassl:1996a}, this overhead could be significantly
reduced.

Here we apply the techniques of of quantum error-detection and
correction to greatly increase the efficieny of implementing quantum
information processing by LOQC.  We first show that for ideal LOQC
(where errors in devices are ignored), it is possible to use a two
qubit error-detecting code to rapidly eliminate the probability of
failed implementations of the two qubit gates. This follows from a
threshold analysis demonstrating a threshold of $.5$ for an error
model where the errors are $\sigma_z$ measurements at known locations.
One consequence of this result is that the non-deterministic gates
of~\cite{knill:qc2000b} need only be implemented using the $2$- or
$3$-photon prepared states.  The next task is to demonstrate that the
main sources of device errors can be efficiently eliminated. We
accomplish this by adding phase error-correction, and more importantly
erasure coding methods. The latter serves the purpose of removing
errors caused by particle loss and detector inefficiency, exploiting
the built-in loss detection capabilities of basic LOQC.  In the
process we give a conservative bound on the threshold for the erasure
error model.

On the basis of this paper and~\cite{knill:qc2000b} we can propose a
roadmap for experimental and theoretical work toward implementing
LOQC. It is based on viewing basic LOQC as a fundamental model of
computation that can be used to implement standard quantum
computation by using a set of layered techniques.  For
benchmarking purposes, the relevant resources of LOQC can be taken to
be the number of particles independently generated, the probability of
(detected) failure of the implemented computation (without
post-selection), and the measured error in the output conditional on
success (i.e.  with post-selection).  The roadmap for LOQC based
quantum computation can be outlined as follows:
\begin{itemize}
\item[1.] Basic LOQC.
\begin{itemize}
\item[a.] Non-deterministic non-linear sign changes.
\item[b.] Preparation of the state $\ket{t_n}$ for teleportation
(see~\cite{knill:qc2000b} for the definition of $\ket{t_n}$).
\item[c.] Teleportation using $\ket{t_n}$ for increasing $n$ with
probability of success close to $1/(n+1)$.
\item[d.] Controlled sign changes with good probability of success.
\end{itemize}
\item[2.] QC based on LOQC.
\begin{itemize}
\item[a.] Boosting the success probability by the use of encoding.
\item[b.] Decrease phase errors by applying error correction.
\item[c.] Decrease loss errors by encoding with erasure codes.
\item[d.] Concatenation (or larger codes) for achieving high accuracy.
\end{itemize}
\end{itemize}
The experimental challenges include the ability to use multiple
independently generated single photons and to control the photonic
qubits using feedback from detectors.  The first demonstrations are
likely not to involve feedback, but rather to use post-selection and
high repetition rates to demonstrate success. Although feedback can be
delayed in our schemes, this comes at the cost of high failure rates
in the state preparation protocols, particularly for those states
that depend on previously prepared states. The ultimate goal is to
build the state preparation protocols into state factories with the
ability to attempt failure-prone state preparations at high rates
(perhaps in parallel), exploiting the ease with which photons can be
generated.

Here is the outline of this paper.  We begin by recalling the needed
properties of the LOQC protocols given in~\cite{knill:qc2000b}.  We
briefly describe the concatenation method for establishing thresholds,
give a code suitable for increasing the success probability in ideal
LOQC, and show how to implement encoded operations.  The gain in
success probability is estimated so as to determine a threshold. The
resources that determine the general error propagation behavior are
bounded for the goal of approaching the quantum communication
threshold.  As a result we conservatively estimate that a gate with
the necessary reliability depends on at most a few hundred LOQC controlled
sign flips implemented non-determinstically with a pair of states each
involving three photons.  More are used up in unsuccessful state
preparations, but at least in principle, this overhead is not much
larger.  The quantum coding methods are then enhanced for the purpose
of dealing with phase and detected-loss errors, including particle
detector inefficiencies, and finally for dealing with any residual
general errors.

We assume familiarity with quantum
computation~\cite{aharonov:qc1998a,divincenzo:qc2000a} and quantum
error-correction via stabilizer
codes~\cite{gottesman:qc1996a,gottesman:qc1997a}. For the basic ideas
of LOQC see~\cite{knill:qc2000b}. Most of this paper is written in the
language of qubits using products of Pauli operators.

\section{Features of LOQC}
\label{sect:features_of_loqc}

LOQC is a model of quantum computation where some of the
gates are non-deterministic, with detectable failures.  The
probability of failure of the gates depends on the resources
used. Specifically, the model is characterized by
enabling the following operations on standard qubits (which are
encoded in the physical system as photonic qubits):
\begin{itemize}
\item[1.] Preparation of $\ket{0}$: $P_0$.
\item[2.] Measurement in the basis $\ket{0},\ket{1}$: $M_0$.
\item[3.] Every one qubit rotation.
\item[4.] Controlled sign flip: $\mbox{c-$\sigma_z$}$, with a
probability of failure, which is detected.
\end{itemize}
In practice errors other than detected failures occur.  For the
moment, we assume that such errors are significantly less likely than
detected failures, so that initially, they can be ignored.
Specifically, this leads to designing implementations by dealing with
errors in order of their importance. We call LOQC with only the errors
due to detected failure of the conditional sign flip ``ideal LOQC''
(iLOQC).  In iLOQC, single photon state preparation, particle number
detection and passive linear optical elements are all perfect.  As
explained in our previous paper~\cite{knill:qc2000b}, when applying
$\slb{\mbox{c-$\sigma_z$}}{12}$, with probability $f$, qubit $1$ is
measured in the $\sigma_z$ basis. If this event does not occur then
with probability $f$, qubit $2$ is measured in the $\sigma_z$ basis.
Thus the prior probability of the second event is $f(1-f)$.  The
measurement outcome is known in either case and the qubit not measured
is preserved.  We take that to be the error model of iLOQC.  Note that
the failure behavior is asymmetric, so that the ordering of the labels
is significant. We call the first qubit in this operation the
``source'', and the second the ``target''. The resource overhead for
implementing this gate depends on $f$. If the methods
of~\cite{knill:qc2000b} are used for implementing the
$\mbox{c-$\sigma_z$}$ operation, a state with approximately $(1/f)-1$
photons needs to be adjoined.  The preparation of the state needs to
be tried several times, using ancillary photons.  With the naive
method, the total number of photons used in the preparation attempts
grows exponentially in $1/f$, so it is desirable to show that we can
scale with as high a probability of failure as possible.

For the present purposes, it is convenient to use a gate set based on
the product operator formalism~\cite{sorensen:qc1983a}.  Thus, gates
are exponentials of products of Pauli operators.  To simplify the
notation, define $X\defeq\sigma_x,Y\defeq\sigma_y,Z\defeq\sigma_z$.
For a product operator $U$, write $U_\theta=e^{-iU\pi\theta/360}$ and
note that since $U^2=I$, $U_\theta=\cos(\pi\theta/360)-i\sin(\pi\theta/360)U$.  For
example, $X_{180}$ is a bit flip up to a global phase.  When there is
no possibility for confusion, we abbreviate $UV\doteq
\slb{U}{a}\slb{V}{b}$, where the parenthesized superscripts are system
labels.

One reason for why product operators are convenient is because it is
straightforward to follow their evolution under $90^\circ$ rotations
by using appropriate triples of axes.  For example, a $Y_{90}$
rotation takes the $Z$ axis to the $X$ axis, so that
$(Y_{90})Z(Y_{-90})=X$.  This implies that if in a quantum network, a $Z$
measurement or a $Z$ rotation precedes a $Y_{90}$ rotation, then this is
equivalent to an $X$ measurement or rotation (by the same angle) after
the $Y_{90}$.  A complete set of gates that generates the
determinant $1$ unitary matrices is given by
\begin{itemize}
\item[1.] $X$ rotations: $X_{\phi}$ for any angle $\phi$.
\item[2.] $Z$ rotations: $Z_{90}$.
\item[3.] $ZZ$ rotations: $(\slb{Z}{1}\slb{Z}{2})_{90}$.
\end{itemize}
(With the use of ancillas, $Z_{90}$ may be eliminated from the set
without loss of completeness.)
The third gate is related to $\mbox{c-$\sigma_z$}$ by
\begin{equation}
(\slb{Z}{1}\slb{Z}{2})_{90} =
e^{i\pi/4}\slb{Z_{90}}{1}\slb{Z_{90}}{2}\slb{\mbox{c-$\sigma_z$}}{12}.
\end{equation}
As a result, the $ZZ$ rotation is readily implemented in LOQC up to
global phases. Furthermore, we can modify the implementation to
achieve the following failure behavior for
$(\slb{Z}{1}\slb{Z}{2})_{90}$: With probability $f$, qubit $1$ is
measured in $Z$ and qubit $2$ is untouched.  With probability
$(1-f)f$, qubit $2$ is measured in $Z$.  If the outcome is $0$, qubit
$1$ experienced a $Z_{90}$ and if it is $1$, a $Z_{-90}$. Note that
due to the availability of $X_{90}$ rotations, we can also use
$(\slb{U}{1}\slb{V}{2})_{\pm 90}$ rotations for $U$ and $V$ either $Z$
or $Y$ with similar error behavior, where the measurements commute
with the rotation.  Similarly, it is straightforward to implement the
$Y_{\pm 90}$ rotations, $Y$ measurement and $Y$ eigenstate
preparation.  We will find that the encoded $Z$ rotation also has a
probability of failure, where a $Z$ measurement occurs if it fails.
In this case, $X$ eigenstate preparation (using $\ket{0}$ preparation
and an $X_{90}$ followed by a $Z_{\pm 90}$) may fail. However, since
the preparation is successful if failure has not been detected, 
it is possible to retry it until success is achieved. This makes it
possible to get a perfect $X$ eigenstate preparation.

\section{Establishing accuracy thresholds}

For the very special error model introduced in the previous section,
accuracy threshold analyses are much simpler. The basic principle is
to use a quantum code that permits implementing the basic operations
on the encoded information in such a way that the new error model is
consistent with the original one, and so that the new error rate is
substantially less.  The encoded qubits then behave just like the more
fundamental ones used to encode them, so that the same coding method
can be used with these new types of qubits. This recursive coding
method is known as concatenation and has the property that the
error-rates decrease super-exponentially with the number of levels
of concatenation. The basic components of an accuracy threshold result
by concatenation are the following:
\begin{itemize}
\item[1.] A quantum code.
\item[2.] Means of implementing each of the basic operations
on the encoded qubit.
\item[3.] Means of recovering from error, which may be part of 2.
\item[4.] Establishing the error model that applies to the encoded
qubits and calculating a bound on the new error rate.
\end{itemize}

The plan is to show that a two qubit code suffices for establishing a
threshold of $f=.5$ for iLOQC.  After dealing with the errors of
iLOQC, it is necessary to consider the contributions of other sources
of noise, particularly photon loss and phase error. It turns out that
the same family of codes can be used with phase error correction.  We
recall that LOQC comes with an effective leakage detection scheme and
observe that the basic teleportation protocol used in LOQC can be
enhanced to allow detection of loss from particle detector
inefficiency.  As a result, it possible to use erasure code to
eliminate errors from photon loss. We point out that due to the special
nature of the erasure error model, good accuracy thresholds apply and
error rates can be readily improved with relatively simple codes and
few levels of encoding.

\section{A two qubit quantum code}

Let $\ket{+}\defeq(\ket{0}+\ket{1})/\sqrt{2}$ and
$\ket{-}\defeq(\ket{0}-\ket{1})/\sqrt{2}$ be the eigenstates of
$X$. Up to overall scale factors, the associated projection operators
are $I\pm X$. We continue to omit these scale factors in identities.
To improve the failure probabilities, we use a two qubit quantum code
with the encoding
\begin{eqnarray}
\ket{0}&\rightarrow&\ket{0_L}\nonumber\\
 &=& (\ket{++}+\ket{--})/\sqrt{2}\nonumber\\
 &=& (\ket{00}+\ket{11})/\sqrt{2}\\
\ket{1}&\rightarrow&\ket{1_L} \nonumber\\
 &=& (\ket{++}-\ket{--})/\sqrt{2}\nonumber\\
 &=& (\ket{01}+\ket{10})/\sqrt{2},
\end{eqnarray}
and show how to implement the necessary operations on the encoded
states. The encoded states define the state space of the ``logical
qubit''. In the language of stabilizer codes, the code is defined by
the stabilizer $\slb{X}{1}\slb{X}{2}$, and accordingly, the projection
onto the code space is given by $I+\slb{X}{1}\slb{X}{2}$.
We use $L$ as the label for logical (encoded) operators and states.
With this encoding, logical operators are given by
\begin{eqnarray}
\slb{X}{L} &=& \slb{X}{1} =_L \slb{X}{2}\\
\slb{Z}{L} &=& \slb{Z}{1}\slb{Z}{2} =_L -\slb{Y}{1}\slb{Y}{1}\\
\slb{Y}{L} &=& \slb{Y}{1}\slb{Z}{2} =_L \slb{Z}{1}\slb{Y}{2},
\end{eqnarray}
where we introduced the notation $=_L$ to denote
identity when restricted to the code.

For the purposes of establishing a threshold, we
use the following set of basic operations and assumptions:
\begin{itemize}
\item[1.] $X$, $Y$ and $Z$ eigenstate (eigenvalue $1$) preparation:
$P_X, P_Y, P_Z$. 
\item[2.] $Y$ and $Z$ measurements: $M_Y$ and $M_Z$.
$M_Z(s)$ means that the result of the $Z$ measurement
was eigenvalue $s$.
\item[3.] $X_{180}$, $Y_{180}$ and $Z_{180}$ rotations. 
\item[4.] $X_{\phi}$ rotations. 
\item[5.] $Z_{90}$ rotation, with failure probability $f$ after
the first encoding.
\item[6.] $(\slb{Z}{1}\slb{Z}{2})_{90}$ rotation, with failure
probability $f$ in our error model.
\end{itemize}
Operations 1. to 4. are error-free in iLOQC. The $Z_{90}$ rotation is
error free in iLOQC, but as we will see, it may fail with probability
$f$ with a $Z$ measurement after the first level of encoding. Note
that $Y_{90}$, $(YZ)_{90}$ and $(YY)_{90}$ rotations can be implemented by
conjugation with error-free $X$ rotations.

\subsection{State preparation}

To encode an arbitrary state $\kets{\psi}{1}$ to
$\kets{\psi}{L}$, one can use the sequence of
operations given by
\begin{equation}
\slb{E}{12}=\slb{Y}{1}_{90}(\slb{Z}{1}\slb{Y}{2})_{90}\slb{Y}{1}_{-90}\slb{P_Z}{2}.
\end{equation}
To see that this works, follow the effect of the unitary gates on the
initial operators $\slb{X}{1}$, $\slb{Z}{1}$ (basic operators
associated with the state to be encoded) and $I+\slb{Z}{2}$ (the
projection onto the prepared state). It can be seen that
$\slb{X}{1}\rightarrow\slb{X}{1}=_L\slb{X}{L}$, $\slb{Z}{1}\rightarrow
-\slb{Y}{1}\slb{Y}{2}=_L\slb{Z}{L}$ and $I+\slb{Z}{2}\rightarrow
I+\slb{X}{1}\slb{X}{2}$ (the projection onto the code).

The sequence $\slb{E}{12}$ can be used to prepare encoded eigenstates
of $X$, $Y$ or $Z$ by preparing the eigenstate on qubit $1$ first. The
process fails with probability $f+(1-f)f = f(2-f)$, and can be
repeated an expected $1/(1-f(2-f))$ times to successfully prepare it.
Later, it will be the case that the $Y$ rotation in the sequence can
fail, which changes the failure probability to
$f(1+(1-f)+(1-f)^2+(1-f)^3+(1-f)^4)$. The resource usage can be
improved by reusing qubits not affected by the measurement in the
failure.

\subsection{Measurement}

To measure the logical qubit in the logical basis,
measure both of the supporting qubits. A $Z$ measurement
on both yields a logical $Z$ measurement via the total
parity of the two outcomes. Similarly, a $Y$ measurement
on the first qubit and a $Z$ measurement on the second
gives a logical $Y$ measurement. A logical $X$ measurement
is accomplished by measuring $X$ on the first qubit.
All of these measurements are without error.

\subsection{Logical qubit rotations}

The logical $180^\circ$ rotations are implemented
by applying basic ones to each qubit. Thus
\begin{eqnarray}
\slb{X}{L}_{180}&=&\slb{X}{1}_{180}\\
\slb{Z}{L}_{180}&=&\slb{Z}{1}_{180}\slb{Z}{2}_{180}\\
\slb{Y}{L}_{180}&=&\slb{Y}{1}_{180}\slb{Z}{2}_{180}.
\end{eqnarray}
These are error-free.  The ability of implementing $180$'s in this way
is generic for stabilizer codes~\cite{gottesman:qc1997a}.  The
rotations $\slb{X}{L}_{\phi}$ are obtained by applying
$\slb{X}{1}_{\phi}$ and are error-free.  To implement
$\slb{Z}{L}_{90}$, apply $(\slb{Z}{1}\slb{Z}{2})_{90}$.  This is not
error-free, and we show later how to use recovery from $Z$-measurement
to get a smaller probability of failure.

\subsection{$ZZ$ rotation}
\label{sect:zz_rotation}

If the logical qubits are encoded in qubits $1$,$2$ and $3$,$4$,
respectively, the logical $ZZ$ $90^\circ$ rotation can be obtained
by applying $(\slb{Z}{1}\slb{Z}{2}\slb{Z}{3}\slb{Z}{4})_{90}$.
This can be done by the sequence
\begin{equation}
(\slb{Z}{L_1}\slb{Z}{L_2})_{90} = \begin{array}[t]{l}
(\slb{Y}{1}\slb{Z}{2})_{90}(\slb{Y}{1}\slb{Z}{4})_{-90}\\
\ssp(\slb{Z}{1}\slb{Z}{3})_{90}\\
\isp(\slb{Y}{1}\slb{Z}{4})_{90}(\slb{Y}{1}\slb{Z}{2})_{-90},
\end{array}
\end{equation}
where we use the convention that the order of application is right to
left within a line and top to bottom for multiple lines.
Unfortunately, this does not readily yield a logical gate with
significantly less error. To do that requires using the teleportation
techniques of~\cite{gottesman:qc1997a,gottesman:qc1999a}.

\section{Robust teleportation}

\subsection{Basic teleportation}

The basic teleportation protocol transfers an arbitrary
state from qubit $1$ to qubit $3$ by first preparing a state
on qubits $2$ and $3$, then making a measurement on qubits
$1$ and $2$, and finally correcting qubit $3$ by applying
one of the $180^\circ$ rotations. Here is a sequence $\slb{T}{123}$
for a variant of the usual protocol that has better
error behavior for our purposes.
\begin{equation}
\slb{T}{123} = \begin{array}[t]{l}
  \slb{P_Z}{2}\slb{P_Y}{3}\\
\ssp
  (\slb{Y}{2}\slb{Z}{3})_{90}\\
\isp
  (\slb{Z}{1}\slb{Y}{2})_{90}\\
\isp
  \slb{M_Z}{2}(s_1)\slb{M_Y}{1}(s_2)\\
\isp
  (\slb{U(s_1,s_2)}{3})_{180}
\end{array}
\end{equation}
The source in the second coupling evolution is chosen
to be qubit $2$. The necessary correction $U(s_1,s_2)$
can be derived by determining the effect of the
process on an input operator. Using the projection
operators $I\pm Z$ and $I\pm Y$ for the prepared states
and for the effects of the measurement the transformation
for an initial $\slb{Z}{1}$ operator is
\begin{eqnarray}
\slb{Z}{1}(I+\slb{Z}{2})(I+\slb{Y}{3}) &\rightarrow&
(I+s_2\slb{Y}{1})(I+s_1\slb{Z}{2})\nonumber\\
&&\hspace*{-1in}\times
\slb{Z}{1}(I-\slb{Z}{1}\slb{Z}{2}\slb{Z}{3})
          (I-\slb{Y}{2}\slb{X}{3})\nonumber\\
&&\hspace*{-1in}\times (I+s_1\slb{Z}{2})(I+s_2\slb{Y}{1}).
\end{eqnarray}
Using the rules 
\begin{eqnarray}
(I+\slb{Z}{a}\slb{U}{b})(I+s\slb{Z}{a}) =
  (I+s\slb{Z}{a})(I+s\slb{U}{b})\hspace*{.2in}
\\
(I+s\slb{Z}{a})(I+\slb{Y}{a}\slb{U}{b})(I+s\slb{Z}{a}) =
  (I+s\slb{Z}{a})\hspace*{.2in}
\end{eqnarray}
and their variations (continuing to omit constants in
the identities), this evaluates to
$-s_1\slb{Z}{3}(I+s_2\slb{Y}{1})(I+s_1\slb{Z}{2})$.
Similarly, 
\begin{eqnarray}
\slb{X}{1}(I+\slb{Z}{2})(I+\slb{Y}{3}) &\rightarrow&
(I+s_2\slb{Y}{1})(I+s_1\slb{Z}{2})\nonumber\\
&&\hspace*{-1in}\times
\slb{Y}{1}\slb{Y}{2}(I-\slb{Z}{1}\slb{Z}{2}\slb{Z}{3})
          (I-\slb{Y}{2}\slb{X}{3})\nonumber\\
&&\hspace*{-1in}\times (I+s_1\slb{Z}{2})(I+s_2\slb{Y}{1}),
\end{eqnarray}
which evaluates to
to $-s_2\slb{X}{3}(I+s_2\slb{Y}{1})(I+s_1\slb{Z}{2})$.
This implies that
\begin{equation}\begin{array}[b]{rcl}
U(1,1) &=& Y\\
U(1,-1) &=& X\\
U(-1,1) &=& Z\\
U(-1,-1) &=& I\\
\end{array}.\end{equation}
For a nice group theoretic treatment of teleportation,
see~\cite{braunstein:qc2000a}.

The state obtained on qubits $2$ and $3$ before the last rotation
and measurements is a prepared entanglement denoted by
$\kets{t_{e}}{23}$. The idea is to use the built in error-detection
and multiple tries to obtain the state without error.

In order to analyze the propagation of errors, we need to understand
the effect of an unintended $\slb{Z}{1}$ or $\slb{Y}{2}$
measurement before the protocol's end.  Both of these commute with the
two rotations in the protocol.  A $\slb{Y}{2}$ measurement implies
that the net effect of the applied rotations on the third qubit
is a $\slb{Z}{3}_{\pm90}$ (the sign depends on the measurement
outcome). Nothing happens to the first qubit. If a $\slb{Z}{1}$
measurement happens, this directly applies to the first qubit.  The
second qubit experiences a $\slb{Y}{2}_{\pm90}$ rotation.  To simplify
matters we intentionally perform the $Y$ measurement on this
qubit to return to the first case as far as qubit $3$ is concerned.

\subsection{Logical $ZZ$ rotations by teleportation}

One method for implementing the logical $ZZ$ rotation on qubits
encoded in qubits $1,2$ and $3,4$ respectively is to first teleport
the four qubits, then apply the rotation $R_Z =
(\slb{Z}{1}\slb{Z}{2}\slb{Z}{3}\slb{Z}{4})_{90}$.  Since the final
step of the teleportation protocol involves a number of $180^\circ$
rotations, and $R_Z$ is in the normalizer of the Pauli group, we can
instead apply $R_Z$ to the destination qubits in the four copies of
$\ket{t_{e}}$ and apply appropriately modified $180^\circ$ rotations
after the teleportation measurement.  Actually, it is better to apply
$R_Z$ first, use the original $180^\circ$ corrections and note that
the overall effect is equivalent to $R_Z$ or $R_Z^\dagger$ after
teleportation. Which one actually occurred can be determined from the
measurement outcomes. To go from $R_Z^\dagger$ to $R_Z$ it is
sufficient to apply $Z_{180}$'s to each qubit.  Because of the ability
to retry the state preparation until it succeeds, this reduces the
problem of reliably implementing $R_Z$ to the problem of reliable
teleportation.

\subsection{Error recovery by teleportation}

Our methods are designed so that the only error from which
it is necessary to recover is a $Z$ measurement with known outcome
on one of the qubits. It is desirable to implement the recovery
so that at worst, a $Z$ measurement occurs on the logical qubit.
By symmetry, it is sufficient to consider a $Z$ measurement
on qubit $1$. The effect of the measurement is to
apply $(I+s\slb{Z}{1})$, where the sign $s$ is known.
One way of restoring the encoded qubit is to notice
that
\begin{equation}
(I+s\slb{Z}{1})(I+\slb{X}{1}\slb{X}{2})
 = (I+si\slb{Y}{1}\slb{X}{2})(I+\slb{X}{1}\slb{X}{2}).
\end{equation}
Since the first operator is a $90^\circ$ rotation around
$\slb{Y}{1}\slb{X}{2}$, it can be undone by applying its inverse.

A method more easily made reliable is based on the syndrome
measurement technique of error correction.  From this perspective, the
unintended $Z$ measurement creates a superposition of states with two
syndromes, that is eigenvalues of $\slb{X}{1}\slb{X}{2}$.  The encoded
qubit can be recovered by measuring
$\slb{S}{12}\doteq\slb{X}{1}\slb{X}{2}$ and if the eigenvalue is $-1$,
applying $\slb{Z}{1}$. To measure $\slb{S}{12}$ we again use
teleportation, reducing the measurement problem to a state preparation
and teleportation problem (see~\cite{steane:qc1999a} for similar ideas
used to solve the more difficult problem of correcting unknown
errors).  The idea is to measure $\slb{S}{12}$ on the destination
qubits of two copies of $\ket{t_{e}}$ before completing the protocol,
and then infer the eigenvalue from the combination of all measurement
outcomes. The correction operations $U$ of the protocol are
unchanged. To see how this works, implement the protocol by
teleporting qubit $1$ with $\kets{t_{e}}{36}$ and qubit $2$ with
$\kets{t_{e}}{47}$, so that the teleported state ends up in qubits $6$
and $7$. The measurement on qubits $6$ and $7$ can be implemented by
the sequence
\begin{equation}
(\slb{Y}{6}\slb{X}{7})_{-90}\slb{M_Z}{6}(s)(\slb{Y}{6}\slb{X}{7})_{90}.
\end{equation}
Note that the two $X$ operators on qubit $7$ occurring in the
two-qubit rotations can be obtained by conjugating $Z$ operators by a
$Y_{90}$ rotation. If the measurement of $XX$ results in $s=-1$, we
correct the state by applying $\slb{Y}{6}_{180}\slb{Z}{3}_{180}$.  (To
see that this correction works, examine the quantum network, moving
the correction operator back to the beginning by appropriately
changing orientations of rotations by anti-commuting operators, and
then absorbing it at the commuting state preparation steps.)  Again,
we can retry this state preparation until it succeeds. The prepared
state is now given by
$(I+\slb{S}{67})\kets{t_{e}}{36}\kets{t_{e}}{47}$, which is already a
logical qubit state on bits $6$ and $7$.  Suppose teleportation
concludes successfully, with correction operators
$\slb{U(s_1,s_2)}{6}$ and $\slb{U(s_3,s_4)}{7}$.  The resulting state
is the same as if we had applied $(I+s'\slb{S}{67})$ after the
protocol, where $s'$ depends on the $s_i$ ($180^\circ$ rotations only
change the sign of Pauli operators when changing the order of events
in a sequence).  If the state to be teleported is in the code, then
$s'=-1$ is impossible (as the projection is otherwise orthogonal to
the state, resulting in a zero probability event), while if it is in
the $-1$ eigenvalue space of $\slb{S}{12}$, then $s'=1$ is impossible.
Because of the relationship of $s'$ to the $s_i$, which event occurred
can be determined from the combination of $s_i$'s that resulted from
the teleportation measurement, and the necessary correction can be
applied.

\section{Error analysis}

We begin by considering errors that occur in the encoded operations
and how one should respond to such errors.

\subsection{Errors in recovery}
\label{sect:errors_in_recovery}

The recovery procedure is applied when an unintended $Z$ measurement
occurs. Suppose this occurs at qubit $1$. Since the state of qubit $1$
is now known, we can take advantage of this to prepare a state with
the first teleportation step (that is the one involving qubit $1$)
already completed.  This can be done error-free, given a number of
attempts.  Specifically, the first teleportation protocol can be
replaced by preparing the destination qubit in the appropriate
eigenstate of $Z$, and then applying the $XX$ measurement to this and
the target of the second teleportation before completing the
latter. As before, one can use either outcome of the $XX$ measurement,
in this case by applying only a $Z_{180}$ correction, if necessary.

If the teleportation of the second qubit fails at the source of the
relevant rotation, the second qubit is untouched, and we try again.
If it fails at the target, then the second qubit is measured in $Z$,
which implies that the logical qubit is measured in $Z$.  Let $F_r$ be
the probability of failure. By following the different possible
outcomes in the attempts, we get
\begin{eqnarray}
F_r &=& fF_r+(1-f)f\\
\therefore\;\; F_r &=& f.
\end{eqnarray}

\subsection{Errors in the implementation of $\slb{Z}{L}_{90}$}

When applying $(\slb{Z}{1}\slb{Z}{2})_{90}$ to the qubits, the
following can happen: 1. The first qubit is measured in the $Z$
basis. In this case, apply the recovery procedure and if it succeeds,
attempt the operation again. If it fails, the logical qubit is
measured in $Z$.  2. The second qubit is measured in the $Z$ basis
with outcome $s$ and the first qubit experiences a $\slb{Z}{1}_{s90}$.
The effect is the same as if a $\slb{Z}{L}_{90}$ had been applied
before the measurement, so if the subsequent recovery procedure
succeeds, then the desired operation has been applied. Otherwise, the
logical qubit has been measured in $Z$.  The probability of failing
once case 2 is entered is $f$.  The probability of entering case 1 is
$f$. By following the re-attempts going through case 1, we obtain the
equation for the probability of failure
\begin{eqnarray}
F_Z &=& (1-f)f^2 + f^2 + f(1-f)F_Z\\
\therefore\;\; F_Z &=& f^2(2-f)/(1-f(1-f)).
\label{eq:F_Z}
\end{eqnarray}

\subsection{Errors in the implementation of $(\slb{Z}{L_1}\slb{Z}{L_2})_{90}$}

To avoid having to retry the operation we modify the protocol for
implementing the logical $ZZ$ rotation slightly.  Instead of preparing
$4$ copies of $\ket{t_{e}}$ with the $ZZZZ$ rotation already applied,
prepare $k\geq 4$ such copies.  In the end, the destination qubits of
the unused copies are measured in $Z$, so that the effective applied
rotation becomes $(\slb{Z}{L_1}\slb{Z}{L_2})_{\pm90}$, where the sign
depends on the measurement outcomes. Compensate for a minus sign by
applying $Z_{180}$'s to each qubit.  Let the qubits be encoded in
qubits $1,2$ and $3,4$ respectively. Perform the teleportation
protocols for the qubits in this order. When a protocol fails at the
source of the critical rotation, try again with the next available
pair of qubits in the prepared state.  When the protocol fails at the
target, the procedure depends on whether it is the first or second
member of a pair of encoding qubits.  If it is the first, attempt
recovery using the second qubit as usual.  If it is the second, do the
same, but using the already teleported qubit as the first member. In
this case, we do not need to re-attempt teleportation for implementing
the rotation, as in the case of the logical $Z$ rotation. Computing
the probability of failure that the procedure fails by the first
logical qubit being measured gives
\begin{eqnarray}
F_{ZZ} &=& f^2 + (1-f)f^2 +f(1-f)F_{ZZ}\\
\therefore\;\; F_{ZZ} &=& F_Z.
\end{eqnarray}
The probabilities for the second logical qubit being measured 
given successful completion of the first two steps is
the same, as required by 
the assumptions of the model.

\subsection{The threshold}

The threshold $T_d$ for obtaining an improvement in the failure
parameter can be determined by solving $F_{ZZ} = f$, which
gives
\begin{equation}
T_d = .5.
\end{equation}

\section{Resource analysis}

Resource analysis can be used to estimate the effect of residual
errors (not fitting the error model) in the basic operations and to
determine the total overhead of implementing an accurate quantum gate
for standard quantum computation or communication.  The total overhead
includes the expected number of attempts required to prepare the
requisite states. In an actual system, the state preparation attempts
can be arbitrarily parallelized and implemented in independent
high-throughput state factories.  In the system as proposed here, the
states are relatively simple in terms of the lower level
implementation, and success probabilities are reasonable. An explicit
total resource analysis is left as a problem for future work. For now,
our primary concern is how general errors can propagate from the
physical implementation to the encoded qubits. This depends only on
the operations that directly contribute toward the state used in the
final gate via their errors conditional on success.  This property can
be exploited to largely eliminate the problem of inefficient
detectors, see Sect.~\ref{sect:loss_of_particles}.

The analysis that follows is intended as an example for how this can
be done and is completed with an explicit example.  We begin by
counting resources in terms of the operations of iLOQC, counting
separately first the error-free one qubit rotations, state
preparations and measurements in the $Z$ or $Y$ basis ($R_0$), second
the $90^\circ$ $Z$ and $Y$ rotations ($R_1$), and third the $90^\circ$
two qubit rotations with $Y$ or $Z$ operators ($R_2$). This separation
helps with the resource estimate due to the fact that they differ in
resource requirements at the first and later levels. For our purposes
two or three levels are expected to suffice. Note that we are not
counting steps that are required to temporarily store a qubit.
This is necessary in principle, as imperfect memories
without parallelism imply that truly scalable quantum or classical
computing is impossible~\cite{aharonov:qc1996b}. In particular, it is
beneficial to parallelize implementations as much as possible.

As we proceed, we will comment on the expected number of tries for
state preparations. For this purpose, define $q$ as the probability of
failure of the one qubit rotations contributing to $R_1$ and
let $p=1-f(2-f)$ be the total failure probability of the two qubit
rotations contributing to $R_2$.  In a total resource analysis, one
can exploit the fact that $q=0$ at the first level.

\subsection{Resources for teleportation}

The preparation of $\ket{t_{e}}$ requires 
\begin{eqnarray}
R_0(t_e) &=& 2\\
R_1(t_e) &=& 0\\
R_2(t_e) &=& 1.
\end{eqnarray}
Since the probability of success is $(1-p)$, the expected number of
attempts to assure success is $1/(1-p)$.

To prepare $2k$ copies of $\ket{t_{e}}$ with the $(ZZZZ\ldots)_{90}$
rotation applied to the targets using the method given in
Sect.~\ref{sect:zz_rotation} requires
\begin{eqnarray}
R_0(Z^{2k}) &=& 2kR_0(t_e) = 4k
\\
R_1(Z^{2k}) &=& 2kR_1(t_e) = 0
\\
R_2(Z^{2k}) &=& 2kR_2(t_e)+4k+1 = 6k+1.
\end{eqnarray}
The probability of successful preparation is only $(1-p)^{4k+1}$, so
the preparation method needs to be improved.  An efficient (in $k$)
scheme is based on the idea of using a parity containing ancilla to
kick back the desired rotation, and to generate this ancilla in a tree
like fashion~\cite{cmoore:qc1998a}.  The sequence is defined
recursively by: $\slb{S_1}{ab}$ consists of preparing
$\kets{t_{e}}{a}$ and $\kets{0}{b}$, then applying
$\mbox{c-$\sigma_x$}$ from the target of $\kets{t_{e}}{a}$ to
$\kets{0}{b}$. The qubit $b$ is the ``parity''
qubit. $\slb{S_{l+1}}{abd}$ applies $\slb{\mbox{c-$\sigma_x$}}{bd}$ to
the outputs of $\slb{S_{l}}{ab}$ and $\slb{S_{l}}{cd}$, then measures
parity qubit $b$ in the $X$ basis by applying $Y_{90}$ and then
measuring $Z$. If the outcome is $-1$, apply a $Z_{180}$ to each of
the target qubits of the $\ket{t_{e}}$ that make up $a$.  The new
parity qubit is $d$. The controlled-not operation
$\slb{\mbox{c-$\sigma_x$}}{ab}$ is applied with
\begin{equation}
\slb{Z}{a}_{-90}\slb{X}{b}_{90}\slb{Y}{b}_{-90}(\slb{Z}{a}\slb{Z}{b})_{90}\slb{Y}{b}_{90}.
\end{equation}
To kick back the desired rotation to $2^l$ copies of $\ket{t_{e}}$
after $S_{l}$ has been successfully completed, apply $Z_{90}$ to the
parity qubit and measure it in the $X$ basis, applying a $Z_{180}$
correction as before, if necessary.  The last step can be deferred
until after the teleportation when using this for implementing the
logical $ZZ$ operation. The failure response of the algorithm can be
optimized by recovering states as much as possible.  For simplicity,
we assume that the state associated with a failed $S_{l}$ is
discarded.

The resources required for implementing $S_{l}$ can be determined
recursively. 
\begin{eqnarray}
R_0(S_1) &=& R_0(t_e)+2 = 4
\\
R_1(S_1) &=& R_1(t_e)+3 = 3
\\
R_2(S_1) &=& R_2(t_e)+1 = 2
\\
R_0(S_{l+1}) &=& 2R_0(S_l)+2+2^{l-1}
\\
R_1(S_{l+1}) &=& 2R_1(S_l)+4
\\
R_2(S_{l+1}) &=& 2R_2(S_l)+1,
\end{eqnarray}
which one can solve to obtain
\begin{eqnarray}
R_0(S_l) &=& 8\times 2^{l-1}-4
\\
R_1(S_l) &=& 7\times 2^{l-1}-4
\\
R_2(S_l) &=& 3\times 2^{l-1}-1.
\end{eqnarray}
The probability of success of $S_{l+1}$ using
two independent outputs of $S_l$ is
$(1-q)^4(1-p)$, which can be shown to  imply
polynomial total resource use.
We now obtain new expressions for
the $Z^k$ preparation resources (assuming that the
last correcting series of $Z_{180}$'s is deferred):
\begin{eqnarray}
R_0(Z^{2^l}) &=& R_0(S_l)+1 = 7\times 2^{l-1}-2
\\
R_1(Z^{2^l}) &=& R_1(S_l)+2 = 7\times 2^{l-1}-2
\\
R_2(Z^{2^l}) &=& R_2(S_l) = 3\times 2^{l-1}-1.
\end{eqnarray}
The success probability given the output of $S_l$
is $(1-q)^2$.

To prepare the state needed for recovery after one of the qubits has
been measured, follow the part of the protocol that does not involve
the remaining qubit. There are two measurements that need to be made, and
we note that the state can be used for completing the protocol
regardless of the outcome.  As explained in
Sect.~\ref{sect:errors_in_recovery}, the preparation can be decomposed
into making a copy of $\ket{t_{e}}$ and of an eigenstate of $Z$ and
then measuring $XX$, correcting the outcome
with a $Z_{180}$ if ncessary. Using the implementation of the $XX$
measurement above and counting the two $Y$ rotations
needed to obtain the $X$ operators from $Z$'s in the couplings, we get
\begin{eqnarray}
R_0(r_e) &=& R_0(t_e)+3=5
\\
R_1(r_e) &=& R_1(t_e)+2=2
\\
R_2(r_e) &=& R_2(t_e)+2=3
.
\end{eqnarray}
The success probability is $(1-q)^2(1-p)^2$.

\subsection{Resources for operations}

To follow the resource usage through several levels of
concatenation, it is necessary to determine the maximum
resource usage for each category of operations.
First are the one qubit $180^\circ$ rotations,
state preparations and measurements in the $Z$ or $Y$ basis.
Of these, state preparation has the highest resource requirements
for $R_1$ and $R_2$. $R_0$ is highest for the $180^\circ$
rotations. This gives the following estimates:
\begin{eqnarray}
R_0(0) &=& 2
\\
R_1(0) &=& 3
\\
R_2(0) &=& 1.
\end{eqnarray}
The next category consists of the $90^\circ$ $Z$ or $Y$ rotations.
For the $Y$ rotation, we conjugate a $Z$ rotation by logical
$X_{90}$'s. The resource analysis is complicated by
the need for using recovery operations on partial failures.
The expected number of $ZZ$ rotations
that need to be retried after failure and successful
recovery of the first qubit is $1/(1-f(1-f))$, as $f(1-f)$
is the probability of failing and then successfully recovering.
To get a better bound on the number of directly contributing
operations, note that the $ZZ$ couplings are implemented
in such a way that if the source fails, the target is
not touched. Thus teleportation steps that fail at the
source of the coupling that precedes the measurements
need not be counted except when estimating total resources.
Thus, the expected number of contributing
recovery operations can be bounded by
$1/(1-f(1-f)) - 1 = f(1-f)/(1-f(1-f))$ for failures at the first qubit
and $f(1-f)(1-f^2/(1-f(1-f))) = f(1-f)^2/(1-f(1-f))$
for failures
at the second qubit. Thus
\begin{eqnarray}
R_0(1) &=& 2+f(1-f)(2-f)/(1-f(1-f)) R_0(r_e)
\nonumber\\
&\leq& 2 + 5 f(1-f)(2-f)/(1-f(1-f))
\\
R_1(1) &=&f(1-f)(2-f)/(1-f(1-f))) R_1(r_e)
\nonumber\\
&\leq& 2f(1-f)(2-f)/(1-f(1-f))
\\
R_2(1) &=& 1/(1-f(1-f)) \nonumber\\&&\mbox{}+ f(1-f)(2-f)/(1-f(1-f))) R_2(r_e)
\nonumber\\
&\leq& 1/(1-f(1-f)) \nonumber\\&&\mbox{}+ 3f(1-f)(2-f)/(1-f(1-f)).
\end{eqnarray}
The final category has the $90^\circ$ couplings.
Except for at most $4$ $X_{90}$ rotations needed
to get $Y$ operators, it suffices to determine
the requirements for the logical $ZZ$ operation.
Let $2^l$ be the number of copies of $\ket{t_e}$ used
in the prepared state. The calculation
is similar to that for $R_x(1)$. Noting that
the probability of a teleportation failing
at the target and the subsequent recovery succeeding
is $f(1-f)^2$, one can bound the expected number of
recovery attempts by $4f(1-f)^2/(1-f(1-f)^2)$.
Some of the teleportations are attempted multiple
times (just like the rotation is attempted
multiple times in the previous case) and we need to account
for the correction steps of the teleportation.
\begin{eqnarray}
R_0(2) &=& 8+2/(1-f(1-f))+R_0(S_l)\nonumber\\&&\mbox{}+(4f(1-f)^2/(1-f(1-f)^2))R_0(r_e)
\nonumber\\
&\leq&8\times 2/(1-f(1-f))+2^{l-1}\nonumber\\&&\mbox{}+20f(1-f)^2/(1-f(1-f)^2)+4\nonumber\\
\\
R_1(2) &=&R_1(S_l)\nonumber\\&&\mbox{}+(4f(1-f)^2/(1-f(1-f)^2))R_1(r_e)
\nonumber\\
&\leq& 2^{l-1}+(8f(1-f)^2/(1-f(1-f)^2))-2\nonumber\\
\\
R_2(2) &=&1/(1-f(1-f))+R_2(S_l)\nonumber\\&&\mbox{}+(4f(1-f)^2/(1-f(1-f)^2)) R_2(r_e)
\nonumber\\
&\leq&
1/(1-f(1-f))+3\times 2^{l-1}\nonumber\\&&\mbox{}+12f(1-f)^2/(1-f(1-f)^2)-1.\nonumber\\
\end{eqnarray}

\subsection{An explicit example}

Suppose the goal is to have a $5\%$ failure probability per qubit,
which is not far from the current best estimates for the communication
threshold~\cite{briegel:qc1998a}. If we use pairs of three-photon
entanglements to generate the controlled sign flip at the first level
in LOQC, the initial failure probability per qubit is $f=1/4$. Using
the expression for $F_Z$ in Eq.~\ref{eq:F_Z} gives $f=0.135$ after one
level and $f=0.038$ after two.  For simplicity, we choose $l=3$ in
both levels (hopefully a safe choice, though in principle this effects
the probabilities a bit). Other useful values at the first and second
levels are $f/(1-f) =0.333, 0.156$, $1/(1-f(1-f)) = 1.23, 1.132$ and
$f/(1-f(1-f)) = .308, .152$.  We evaluate the values for the coupling
evolution at the second level.  Using parenthesized superscripts to
denote the levels gives, for example,
\begin{eqnarray}
\slb{R_0}{2}(2)
 &=&
 (7\times 2^2+20\times 0.156+9)\slb{R_0}{1}(0)
\nonumber\\&&\mbox{}+
 (2^2+8\times 0.156-2)\slb{R_0}{1}(1)
\nonumber\\&&\mbox{}+
 (3\times2^2+12\times 0.156 + .5)\slb{R_0}{1}(2)
\nonumber\\
&\leq& 656
\end{eqnarray}
Similarly,
\begin{eqnarray}
\slb{R_1}{2}(2) &\leq& 169
\\
\slb{R_2}{2}(2) &\leq& 239
\end{eqnarray}
Much of the inefficiency comes from the lack of optimization
for implementing the logical $ZZ$ rotation.

The expected total resource usage including those needed for the
failed state preparation attempts can be determined from the
probabilities of successful preparation and requires recalculating the
expressions for the various resources. For example, at the first
level, $p=p_1=0.44$ and $q=q_1=0$ and at the second, $p=p_2=.25$ and
$q=q_2=.135$.  Thus the expected number of attempts required to make
the state needed for error recovery is approximately $3.16$ at the
first level and $2.38$ at the second. 

The resource values imply that a controlled sign flip at the top level
depends on less than $250$ controlled sign flips implemented with
pairs of three photon entanglements in iLOQC. This can be used to
bound the effect of errors that cannot otherwise be controlled.
The resource bounds improve if four or five photon states $\ket{t_n}$
can be reliably generated, thus giving better initial values of $f$
and substantial gains in efficiency at the higher levels.

\section{Compensating for other errors}

So far we have shown how to use a simple code
with careful implementation of the basic operation to
rapidly boost the probability of successful completion
of gates. The next step is to consider the contribution
of other errors and how to correct for them.
Two important types of errors are phase errors and loss
of particles.

\subsection{Phase errors}

The occurrence of a phase error can be detected for the code used
above by following the teleported $XX$ measurement procedure in the
absence of a failure. A $-1$ eigenvalue indicates an error. At this
point, it is necessary to return the state to the code space by
applying an appropriate $180^\circ$ rotation.  Although it is not
possible to correct for the error (we don't know which qubit is
faulty), its detection can be used as information for correction in a
higher level erasure code. An alternative is to use the generalization
of the code to three qubits, which does have the capability of
correcting for phase errors. In fact, the $k$-fold concatenation of
the two qubit code with itself is actually a $2^k$ qubit code that
corrects for up to $2^{k-1}-1$ phase errors.  That property can be
exploited by introducing an appropriate error detection/correction
procedure periodically after the first few levels, without changing
the overall concatenation scheme or the basic methods for implementing
operations.

\subsection{Loss of particles and detector inefficiency}
\label{sect:loss_of_particles}

The methods of LOQC include one that can, with some probability of
failure, detect loss of a photon used to define a photonic qubit.  The
failure mode is again one involving a $Z$ measurement, so the scheme
can be used with the two qubit code.  This will have an effect on the
overall error behavior. We leave the calculations as an open problem.

One observation not made in~\cite{knill:qc2000b} is that if after any
of the teleportation steps used for basic LOQC, we measure the modes
not containing the teleported qubits, and the total number of
photons detected is not equal to the number initially
prepared in the entanglement, then a photon was lost, possibly due to
detector inefficiency. Such an event can be declared as a 
qubit loss. Doing so turns the problem of detector inefficiency into one of
having to handle detected qubit loss at some rate.  It is a useful
task to determine the relationship between detector inefficiency and
the probability of detected loss.

If total loss is detected, the two qubit code is insufficient for
restoring the encoded information.  Instead, it is necessary (perhaps
after a few levels of two qubit encodings) to use erasure codes. These
are codes with the property that one (or a few) lost qubits can be
restored without loss of information. The accuracy threshold for the
error model where each operation satisfies that the target qubits are
lost with independent probability $s$ appears to be very good also,
perhaps below $95\%$. A brief explanation based on conservative
calculations giving a value close to $99\%$ is below.  The erasure
model is one of few for which it is possible to establish exact
quantum communication quantities, such as the quantum channel
capacity~\cite{bennett:qc1997b}.  We therefore suggest that
calculating the threshold for the erasure error-model is an excellent
open problem to solve.

\subsection{An erasure code}

A useful erasure code encoding one qubit into four is
defined by the stabilizer group generated by
\begin{equation}
S_E = \{
\slb{X}{1}\slb{X}{2}\slb{X}{3},
\slb{Y}{2}\slb{Y}{3}\slb{Y}{4},
\slb{Z}{1}\slb{Z}{3}\slb{Z}{4}
\}
\end{equation}
One can define encoded operators by
\begin{eqnarray}
\slb{Z}{L} &=& \slb{Z}{2}\slb{Z}{3}\\
\slb{X}{L} &=& \slb{X}{3}\slb{X}{4}.
\end{eqnarray}
We chose the erasure code and the logical operators for their small
support (number of non-identity Pauli operators in the products).  If
it is desirable to encode two qubits into four, the erasure code with
stabilizer generated by $XXXX$ and $ZZZZ$ can be used.

It turns out that it is easier to analyze thresholds for erasure
errors than for $Z$ measurements.  Implementations of operations can
be based on teleportation the same way we did before, again relying on
the ability to guarantee prepared states by retrying after failure.
For the present discussion, we implement the teleportations required
for an operation all in one step, and then follow up with error
recovery if necessary. Of the operations required for quantum
computing, state preparation can be implemented error-free at all
levels by retrying the process until success.  To measure
$\slb{Z}{L}$, first measure $\slb{Z}{2}$, and if this fails, restore
the logical qubit. After successfully measuring $\slb{Z}{2}$, measure
$\slb{Z}{3}$, and if that fails, $\slb{Z}{1}$ and $\slb{Z}{4}$.  The
value of the $\slb{Z}{3}$ measurement can be inferred by using the
parity constraint associated with the third generator of $S_E$.  The
other logical Pauli operators can be measured similarly.  The
probability that the measurement fails is bounded by $sr+3s^2$, where
$r$ is the probability that the error recovery procedure fails.

Suppose that qubit $1$ is lost. The error recovery procedure requires
measuring the first and third operators of $S_E$, which, by
reintroducing a fixed state for qubit $1$, can be done with two full
teleportation steps each. We ignore the fact that some failures in
these steps can be recovered and estimate that each teleportation
step fails with probability at most $3s$ (due to the coupling rotation and
two measurements, not including the correction for now). The recovery
step concludes with a correcting operation on qubit $1$, so that we
can estimate $r\leq 13s$ (ignoring the possibility of retrying the
recovery step if one of the last corrections fail). The probability
of a failed measurement is therefore bounded by $16s^2$.

Since we are unable to implement arbitrary rotations directly, it is
necessary to implement (say) a $45^\circ$ rotation by teleportation.
(The compensation step is now a $90^\circ$ coupling rotation, which
can be implemented either by teleportation or directly, exploiting the
independence assumption on loss of qubits.)  Since the most complex
operation involves coupling two logical qubits with a $90^\circ$
rotation, we finish by giving a rough estimate of this failure
probability. The Hamiltonians to be evolved have weight four (two on
each pair of encoding qubits), so it suffices to teleport four qubits
after a suitable state preparation. If one of the teleportations fail,
the recovery procedure is followed. We may need to compensate for a
negative rotation induced by the teleportations' last correction step,
but that can be absorbed into the procedure. Thus the error
probability for the first logical qubit is bounded by
$(8s)(13s)=104s^2$, whence an accuracy threshold 
better than about $99\%$.

\subsection{General errors}

The techniques discussed so far compensate for all of the primary
errors that are expected to occur in an LOQC system.  Additional
errors can be attributed to improper settings of beam splitters,
undetected loss, stray photons, etc.  The first of these depends on
accurate calibration of classical control parameters. Happily, this
type of error affects the probabilities quadratically, so that it can
be minimized well by engineering.  To deal with other errors
eventually requires more powerful quantum error correction, and the
goal of any implementation is to minimize the need for these
techniques. 

\section{Conclusion}

This work is a first attempt at reducing the overhead and need for
efficient devices for implementing LOQC.  It shows that even without
much optimization of the operations or tight estimates of errors and
resources, there are techniques that can be used to obtain useful
quantum operations in LOQC with overheads within two orders of
magnitude of those required for other reliable implementations of
quantum computers. The advantages of LOQC include the ability to
compensate readily for the primary errors in optics while preparing
large numbers of the basic quantum systems, which are photons in a
superposition of two modes.  It is necessary to further optimize the
methods and to better analyze the error behavior, particularly the
effects of detector inefficiencies, single photon state preparation
errors and timing or overlap problems for photons.  A fruitful area of
further investigation is to determine whether larger codes and the
method of encoding multiple qubits at once can be used to improve
efficiency and error behavior.

Our proposal consists of a multi-level system with various types of
error-control gradually introduced and supported by high-output state
preparation factories that exploit the ease with which many photons
can be produced.  Although in the very long term, solid state or
molecular computing methods are the preferred implementation for large
scale quantum computing, LOQC is now a viable alternative to achieving
the capability of non-trivial quantum information processing. One
area where LOQC has a long term future is in communication.  Photon
based systems are currently the only reasonable proposals for long
distance quantum communication. Since the necessary accuracies for
successfully exchanging entanglement over arbitrary distances are well
below $99\%$, this may also be the first application of LOQC to
quantum information processing to be experimentally implemented.

Acknowledgments: We thank the Aspen Center for Physics for its
hospitality. E.K. and R.L. received support from the NSA
and from the DOE (contract W-7405-ENG-36).

%\bibliography{journalDefs,qc}

\end{document}